\documentclass[12pt]{iopart}
\usepackage{eurosym}
\usepackage{epsfig}
\usepackage{bm}
\usepackage{amssymb}
\usepackage{amstext}
\usepackage{color}

\begin{document}

\title{Bose-Hubbard model with occupation dependent parameters}
\author{O. Dutta$^1$, A. Eckardt$^1$, P. Hauke$^1$, B. Malomed$^2$, M.
Lewenstein$^{1,3,4}$}
\address{${}^1$ ICFO-Institut de Ci{\`e}ncies Fot{\`o}niques,
Mediterranean Technology Park, E-08860 Castelldefels (Barcelona),
Spain \\$^2$ Department of Physical Electronics, School of
Electrical Engineering, Faculty of Engineering, Tel Aviv University,
Tel Aviv 69978, Israel
\\${}^3$ ICREA-Instituci{\`o} Catalana de Recerca i Estudis Avan\c{c}ats, Lluis Companys 23, E-08010 Barcelona,
Spain \\ ${}^4$ Kavli Institute for Theoretical Physics, Kohn Hall,
University Of California,
Santa Barbara, California 93106-4030} \ead{omjyoti.dutta@icfo.es}
\date{\today}

\begin{abstract}
We study the ground-state properties of ultracold bosons in an optical
lattice in the regime of strong interactions. The system is described by a
non-standard Bose-Hubbard model with both occupation-dependent tunneling and
on-site interaction. We find that for sufficiently strong
coupling the system features a phase-transition from a Mott insulator with
one particle per site to a superfluid of spatially extended particle pairs
living on top of the Mott background -- instead of the usual transition to a superfluid of single particles/holes. 
Increasing the interaction further, a
superfluid of particle pairs localized on a single site (rather than being
extended) on top of the Mott background appears. This happens at the same interaction strength where 
the Mott-insulator phase with 2 particles per site is destroyed
completely by particle-hole fluctuations for arbitrarily small tunneling. In
another regime, characterized by weak interaction, but high occupation
numbers, we observe a dynamical instability in the superfluid excitation
spectrum. The new ground state is a superfluid, forming a 2D slab, localized
along one spatial direction that is spontaneously chosen.
\end{abstract}

\maketitle

\section{Introduction}

Systems of ultracold atoms in optical lattices provide a unique playground
for controlled realizations of many-body physics \cite{LewensteinReview, ib2}.
For sufficiently deep lattices, the kinetics is exhausted by tunneling
processes, and an initially weak interparticle interaction eventually
becomes important with respect to the kinetics, when the lattice is ramped
up. A consequence of this competition is the quantum phase transition from a
superfluid of delocalized bosons to a Mott insulator where the particles are
localized at minima of the lattice by a repulsive contact interaction \cite%
{fish}. This effect has been observed in seminal experiments with ultracold
rubidium atoms in a cubic lattice \cite{ib1}. It is described quantitatively
by means of the simple Bose-Hubbard model \cite{fish, jak}, whose parameters
are the interaction energy $U$ for each pair of particles occupying the same
lattice site, and the matrix element $J$ for tunneling between neighboring
sites. Intriguing Hubbard-type physics can
also be observed if the above scenario is extended to fermions, mixtures of
several particle species, exotic lattice geometries, or long-ranged dipolar
interaction \cite{LewensteinReview, ib2, lwrev}.

In this paper, we consider a different type of extension of the bosonic Hubbard
model, becoming relevant when the interaction between the particles is enhanced,
e.g, by means of a Feshbach resonance. As long as the interaction is weak
compared to the lattice potential, a system of ultracold atoms can be described,
to a good approximation, in terms of the lowest-band single-particle Bloch or
Wannier states, the latter being localized at the minima of the lattice
\cite{AshcroftMermin}. Under these conditions, the Hubbard interaction $U$ and
tunneling parameter $J$ are given by respective matrix elements with respect
to the single-particle Wannier states. This approximation
corresponds to degenerate perturbation theory up to first order with respect to the interaction, with only the intraband coupling induced by
the interaction taken into account. However, if the interaction is stronger,
higher-order corrections start playing a role. One may still describe the
system in terms of lattice-site occupation numbers $n_{j}$, but the occupied
Wannier-like orbitals will have admixtures from higher bands, depending on
the occupation. The most significant effect of the repulsive interaction
will be a broadening of the Wannier-like orbitals with increasing
occupation, effectively enhancing $J$ and decreasing $U$. In terms of the
Hubbard description, we take this into account by replacing $J$ and $U$ by
functions $J_{\hat{n}_{i},\hat{n}_{j}}$ and $U_{\hat{n}_{i}}$ of the
number operators $\hat{n}_{i}$. Quantitative consequences of this kind of
modification to the plain bosonic Hubbard model have been studied by
several authors at a theoretical level \cite{cjw, niu, mu}. Considering an
interaction-induced modification of the Wannier functions, also additional
Mott-insulator phases have been predicted \cite{ceder,ceder1}.
In Ref. \cite{lars}, the effect of the interaction-induced coupling to the first
excited band on the Mott transition was considered. Re-entrant behaviour in the superfluid-Mott
transition has also been predicted due to the interaction-induced modification of Hubbard parameters \cite{lars, lundh}.
The effect of interaction on the tunneling dynamics in one-dimensional double-well and triple-well potentials have been studied in 
Refs. \cite{sch1, sch2} where the authors found enhanced correlated pair tunneling near the fermionization limit.
Moreover, occupation-number-dependent on-site interaction has been
observed experimentally in the coherent dynamics of an atomic ensemble \cite%
{ib4}. Similar occupation-dependent effects have been observed in Bose-Bose
\cite{Catani2008} and Fermi-Bose mixtures \cite{Ospelkaus2006, Guenter2006, will},
and -- in the latter case -- have been explained theoretically in terms of
occupation-dependent parameters $U$ and $J$ \cite{Luehmann2008}.

In the present work, we show that new quantum phases can arise in Hubbard
models with number-dependent parameters. After writing down the effective
single-band Hamiltonian including the effect of the site occupation, we find
that for strong enough interaction (characterized by the $s$-wave scattering
length $a_{s}$), there is a transition from a Mott state with one particle
localized at each lattice site to a superfluid of \emph{pairs} extended over
neighboring sites, rather than to a superfluid of single atoms. This feature
is novel, considering the fact that the extended pairs emerge in the
single-species repulsive bosonic system without the presence of any
long-range interaction. For even higher interaction strengths, the $n=1$
Mott state becomes unstable towards a superfluid of pairs that are localized
on single sites. Moreover, the $n=2$ Mott state becomes unstable towards
pair fluctuations already for very low tunneling amplitudes. Finally, we
consider the regime where interaction effects are important not because of
large scattering lengths, but rather because of large site-occupations
numbers. In this limit, starting from the Bogoliubov approach to the
homogeneous system, we find a phonon instability at a critical filling
fraction. Above that fraction, the new ground state is a Bose condensate
with the particle density being localized along one spatial direction that
is chosen spontaneously.

This paper is arranged in the following way: In section \ref{bhmodel}, we
introduce the occupation-dependent Bose-Hubbard model. In
section \ref{sip}, we start discussing the properties of this model. Namely, we
study the instability of the Mott-insulator phase with respect to simple
particle and hole excitations, leading to the usual single-particle
superfluidity. In section \ref{eps}, we then investigate the instability
of the Mott phase with respect to the excitation of bond-centered pairs of
particles being extended over neighboring lattice sites. We show that this
mechanism will eventually become relevant when the \textit{s}-wave scattering
length is increased, and that one finds a phase transition to a superfluid of
extended pairs. In section \ref{lps}, proceeding to even stronger interaction,
the instability of the Mott phase towards a superfluid of site-centered pairs
is discussed. In this regime, moreover, the Mott insulator at a
filling of two particles per site can disappear completely. Finally,
in section \ref{bog}, we focus on the limit where interaction-induced orbital
effects play an important role because of large filling. We find that, with the
increasing superfluid density, the condensate may become dynamically
unstable.

\section{The Bose-Hubbard model}

\label{bhmodel} The Hamiltonian in the presence of a periodic potential with
lattice constant $a$, given by $V_{\mathrm{per}}(\vec{r})=V_0[\sin^2(\pi
x/a)+\sin^2(\pi y/a)+\sin^2(\pi z/a)]$, reads
\begin{equation}  \label{oham}
H=\int d^3r \hat{\psi}^{\dagger}(\vec{r}) \left [-\frac{\hbar^2}{2m}
\nabla^2 + V_{\mathrm{per}}(\vec{r}) +\frac{g}{2}|\hat{\psi}(\vec{r})|^2
\right ] \hat{\psi}(\vec{r}),
\end{equation}
with bosonic field operators $\hat{\psi}$, mass $m$, and interaction
strength $g=4\pi\hbar^2 a_s/m$, where $a_s$ is the $s$-wave scattering
length. To derive a Hubbard-type description, the field operators
$\hat{\psi}(\vec{r})$ are expanded in terms of Wannier-like orbitals
$\phi_i(\vec{r},\hat{n}_i)=\phi(\vec{r}-\vec{R}_i,\hat{n}_i)$
localized at the lattice minima
$\vec{R}_i$, namely $\hat{\psi}(\vec{r})
=\sum_i \hat{b}_i \phi(\vec{r}-\vec{R}_i;\hat{n}_i)$ with bosonic
annihilation and number operators $\hat{b}_i$ and
$\hat{n}_i=\hat{b}^\dag_i\hat{b}_i$. Note that the ``wave function''
$\phi_i$ depends on the number operator $\hat{n}_i$ in order to take into
account interaction-induced occupation-dependent broadening. Keeping only
on-site interaction, we arrive at the effective single-band Hamiltonian
\begin{equation}  \label{hub2}
H=-\sum_{ij} J_{\hat{n}_i,\hat{n}_j}\hat{b}^{\dagger}_ib_j+ \frac{1}{2}%
\sum_i U_{\hat{n}_i} \hat{n}_i(\hat{n}_i-1)-\sum \mu \hat{n}_i,
\end{equation}
where
\begin{eqnarray}
J_{\hat{n}_i,\hat{n}_j}&=&-\int d^3r \phi(\vec{r}-\vec{R}_i;\hat{n}_i)
\Big[-\frac{\hbar^2}{2m} \nabla^2   
+V_{\mathrm{per}}(\vec{r}) \Big] \phi(\vec{r}-\vec{R}_j;\hat{n}%
_j+1),  \nonumber \\
U_{\hat{n}_i}&=&g \int d^3r \phi^2(\vec{r}-{R}_i;\hat{n}_i)\phi^2(\vec{r}-%
\vec{R}_i;\hat{n}_i-1),
\end{eqnarray}
and we have introduced the chemical potential $\mu$ to control the particle
number. 
We would like to mention that in the presence of an optical lattice for 
high interactions the pseudo-potential form of contact interaction can still be used, 
when a modified scattering length which is different from the bare scattering length is applied \cite{busch, bij,cui, buch}.

In order to estimate the occupation number dependence in a mean-field way,
we make a Gaussian ansatz for the Wannier-like wave functions,
$\phi (\vec{r} -\vec{R}_{i};n_{i})=\exp (-(\vec{r}-\vec{R})^{2}/d^{2}(n_{i}))$,
where the width $d(n_{i})$ is a variational parameter depending on the particle
number $n_{i}$, and minimize the Gross-Pitaevskii energy functional. The idea to
use the width of the Wannier function as a variational parameter has also
been used in Refs. \cite{chio,vign1,vign2}. Taking into account the full
lattice potential (i.e., not employing a quadratic approximation for the
lattice minima), for a given $n_{i}$ this leads to
\begin{eqnarray}
\left[ \frac{d(n_{i})}{d_{0}}\right] ^{5}\exp \left[ -\pi ^{2}\frac{
d^{2}(n_{i})}{a^{2}}\right] &=&\frac{d(n_{i})}{d_{0}}+\sqrt{2\pi }\left[
\frac{V_{0}}{E_{R}}\right] ^{1/4}\frac{a_{s}}{a}(n_{i}-1).  \nonumber
\label{sig} \\
&&
\end{eqnarray}%
We have introduced $d_{0}/a=\left[ \frac{V_{0}}{E_{R}}\right] ^{-1/4}/\pi $
for the width of $\phi $ in the limit $V_{0}\gg E_{R}$, where $E_{R}=\pi
^{2}\hbar ^{2}/2ma^{2}$ denotes the recoil energy. Note that Eq.~(\ref{sig})
has a solution only as long as $\sqrt{V_{0}/E_{R}}\gg d^{2}(n_{i})/d_{0}^{2}$.
Using the variational result, the tunneling parameter between two adjacent
sites can be approximated by
\begin{equation}
\frac{J_{n_{i},n_{j}}}{E_R}\approx
\left( \frac{\pi ^{2}}{4}-1\right)\frac{V_{0}}{E_{R}}
\exp \left[ -\frac{a^{2}}{2(d^{2}(n_{i}+1)+d^{2}(n_{j}))}\right] .
\end{equation}
We would like to point out that when calculating the tunneling strength, the
Gaussian approximation generally results in a lower value than the exact
calculation; the exact Wannier orbital has an exponential tail which decays
slower than a Gaussian. Nevertheless, our simple approximation provides us with
reasonable numerical values and with a suitable model for the occupation
dependence of tunneling in the regime treated here. This allows us to get a 
qualitative understanding of the physics at work.

For the number-dependent on-site interaction strength the
variational result gives 
\begin{equation}
\frac{U_{n_{i}}}{E_R}=\sqrt{\pi }\left( \frac{V_{0}}{E_{R}}\right)^{3/4}\left[ \frac{%
4d_{0}^{2}}{(d^{2}(n_{i})+d^{2}(n_{i}-1))}\right] ^{3/2}\frac{a_{s}}{a}.
\end{equation}
The single-particle tunneling term arising from the non-on-site contributions of the quartic interaction term in Eq.\ (\ref{oham}) is exponentially smaller than $J(n_i,n_j)$ by approximately a factor of $\exp(-\pi^2\sqrt{V_0/E_R}/4)a_s/a$. Similarly the pair tunneling term is smaller than $J(n_i,n_j)$ by approximately a factor of $\exp(-\pi^2\sqrt{V_0/E_R}/2)a_s/a$. Since we are in the limit of $V_0/E_R \gg 1$, these terms are neglected in Eq.\ (\ref{hub2}).

\section{Insulator to single-particle superfluid transition}

\label{sip} Having written down a suitable model Hamiltonian describing the
regime of strong interaction, we now study the transition from the Mott
insulator having on average $\overline{n}$ particles per site to a superfluid of single particles/holes.

\begin{figure}[ht]
\begin{center}
\epsfig{file=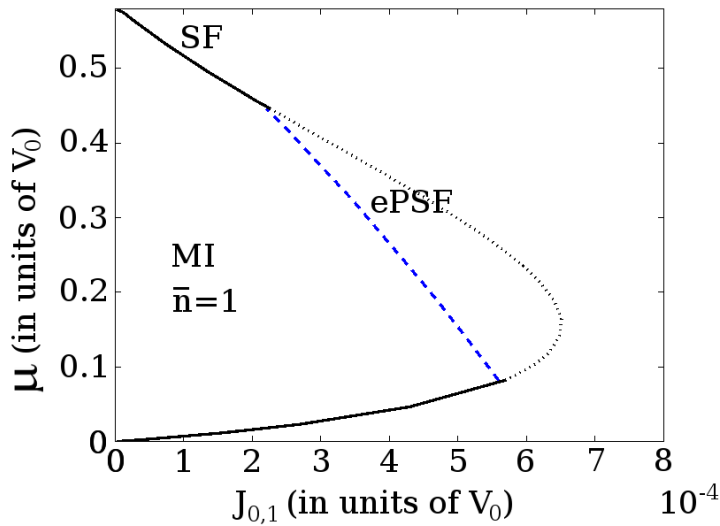,width=10cm}
\end{center}
\caption{Mott-insulator-to-superfluid phase transition for $a_s/a=0.15$.
Inside the region marked by the black solid line and the blue dashed line, the system
is a Mott-insulator with $\overline{n}=1$ particles per site. Leaving this region
by crossing the black solid line, a simple superfluid of single particles
(or, equivalently, holes) is formed (SF).
In contrast, crossing the blue dashed line one arrives at a superfluid phase of extended
(bond-centered) pairs (ePSF). In technical terms of our variational approaches: outside
the black solid and dotted line minimizing the energy (\ref {ven}) gives $\protect\theta\ne0$,
while on the r.h.s.\ of the blue dashed line $\theta_e\ne0$ is obtained from minimizing
expression (\ref{ven1}).}
\label{epp}
\end{figure}

For this purpose, we use a product ansatz $\prod_i |\Phi\rangle_i$ for the
many-body state, with the variational coherent spin-representation state
\cite{aur, blat},
\begin{equation}  \label{sf1}
|\Phi\rangle_i=\cos \theta |\overline{n}\rangle_i + \sin\theta\sin\psi |%
\overline{n}+1\rangle_i+\sin\theta\cos\psi|\overline{n}-1\rangle_i
\end{equation}
at each site $i$, with occupation number basis states $|n_i\rangle_i$. 
Here we only take into account states with one additional particle or hole, which in the Mott phase and close to the transition to the superfluid, where particle fluctuations are small, is sufficient.
Accordingly, 
the variational mean-field energy is given by
\begin{equation}  \label{ven}
\frac{E_{\mathrm{ss}}}{N}=-\frac{zH_J}{4}\sin^22\theta +\left[\frac{H_U}{2}%
+\mu \cos2\psi\right]\sin^2\theta,
\end{equation}
where
\begin{eqnarray}
H_J&=&(\overline{n}^2+\overline{n})J_{\overline{n},\overline{n}}\sin 2 \psi/2
\nonumber \\
&&+ (\overline{n}+1)J_{\overline{n}+1,\overline{n}}\sin^2\psi+\overline{n}J_{%
\overline{n},\overline{n}-1}\cos^2\psi,  \nonumber \\
H_U&=&\overline{n}(\overline{n} - 1)U_{\overline{n}}\cos^2\theta+\overline{n}%
(\overline{n}+1)U_{\overline{n}+1}\sin^2\theta \sin^2\psi  \nonumber \\
&&+ (\overline{n}-1)(\overline{n}-2)U_{\overline{n}-1}\sin^2\theta\cos^2\psi.
\end{eqnarray}
Minimizing the energy determines $\theta$ and $\psi$. While $\theta=0$
corresponds to an incompressible Mott-insulator state with an integer number
of particles $\overline{n}$ per site (found within a finite interval of the
chemical potential $\mu$), the superfluid state is characterized by $%
\theta\neq0$ with order parameter $\langle b_i\rangle\sim \sin 2\theta$. In
the superfluid phase, the average particle number per site is characterized
by $\psi$ depending smoothly on the chemical potential. For $\psi \ll \pi/4$%
, the transition to the superfluid occurs mainly via the creation of
holes, while for $\psi$ near $\pi/2$ particle creation is the main mechanism
destroying the Mott phase. In the latter case, the Mott insulator becomes
unstable when the energy cost of creating
an additional particle at one site, namely $U_{n+1} n(n+1)/2-\mu$, is overcome by the
reduction in energy due to tunneling of that particle, which is on the order
of $z(\overline{n}+1)J_{\overline{n}+1,\overline{n}}$, with coordination
number $z=6$ for the cubic lattice. Thus, when $E_{\mathrm{ss}}$ minimizes
for non-zero $\theta$, the Mott state becomes unstable with respect to
single particle and hole excitations. For interaction strength $a_s/a=0.15$ and $%
\overline{n}=1$ , this happens at the black lines (solid or dotted) in the
plane spanned by $\mu/V_0$ and $J_{0,1}/V_0$ in Fig.~\ref{epp}.

\section{Superfluidity of extended (bond-centered) pairs}\label{eps} 
\begin{figure}[ht]
\begin{center}
\epsfig{file=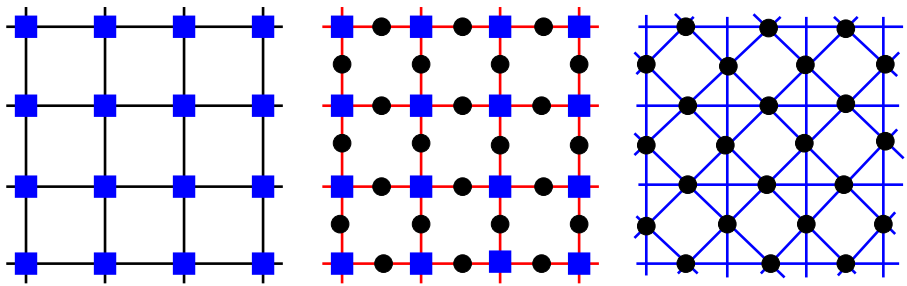, width=12cm}
\end{center}
\caption{Color online: The left hand side shows a square lattice of sites
(blue squares) connected by bonds (black lines). The lattice of the bonds of
the square lattice, where bonds sharing a site are connected, is given by
the checkerboard lattice shown on the right hand side. If a bound pair of
two indistinguishable particles can either occupy a site or a bond of the
cubic lattice (the latter means that the two particles occupy neighboring
sites) and if the pair can move (by single-particle tunneling) from a site
to a neighboring bond and vice versa, then the pairs move on the lattice
shown in the center plot. Sites and bonds are denoted by blue squares and
black bullets, respectively. Extending all the considerations shown in this
figure to the case of a three-dimensional cubic lattice of sites is
straightforward.}
\label{checker}
\end{figure}

So far, we have described the usual scenario of the Mott phase
becoming instable with respect to particle and hole delocalization, as it is
also found for non-number-dependent Hubbard coupling $J$ and $U$. However,
we will now show that---as a consequence of occupation dependent hopping and
on-site interaction---the Mott insulator with $\overline{n}=1$ can become
unstable with respect to the creation of pairs of particles already before
the creation of single particles becomes favorable. Consider a pair
excitation with one additional particle at site $i$ and another one at the
neighboring site $j$, corresponding to the state $|P_{\langle ij\rangle
}\rangle \equiv \frac{1}{2}\hat{b}_{i}^{\dag }\hat{b}_{j}^{\dag
}|\{n_{i}=1\}\rangle $. Such a \emph{bond-centered} or extended pair
excitation at $\langle ij\rangle $ can tunnel coherently to a neighboring
bond, say $\langle ik\rangle $, with $k\neq j$ being another neighbor of $i$%
. Generally, bonds are considered neighbors if they share a common site.
Such a pair tunneling processes occurs in second order with respect to
single-particle tunneling via the virtual site-centered pair state $%
|P_{i}\rangle \equiv \frac{1}{\sqrt{3!}}\hat{b}_{i}^{\dag }\hat{b}_{i}^{\dag
}|\{n_{i}=1\}\rangle $, which has larger energy. According to second-order
degenerate perturbation theory, the amplitude of the pair tunneling process
is given by $J_{\mathrm{eff}}=6J_{2,2}^{2}/(3U_{3}-2U_{2})$. On the same
footing, perturbation theory gives the binding energy of $-2J_{\mathrm{eff}}$
for the bond-centered pair due to number fluctuations within the pair. For a
cubic lattice of sites, the bond-centered pair excitations live on an exotic
lattice of coordination number $z^{\prime }=10$, being a generalization of
the two-dimensional checkerboard lattice (see the rightmost drawing in
Fig.~\ref{checker}) to three dimensions. This allows the pair to reduce its
energy by $10J_{\mathrm{eff}}$ when delocalizing. In contrast, two
additional particles, not forming a pair, can reduce their energy by $%
2\times 6\times 2J_{1,2}$ when delocalizing on the cubic lattice of sites
(coordination number 6). Thus, according to perturbation theory,
the formation of a bond-centered pair is favorable if
$-(10+2)J_{\mathrm{eff}}>24J_{1,2}$.
For certain scattering lengths $a_{s}$, this condition can be fulfilled,
since the Wannier-broadening with increasing scattering lengths leads to an
increase of both $J_{2,2}/J_{1,2}$ and $U_{2}/U_{3}$. In such a situation,
the Mott-insulator state becomes unstable with respect to the creation of
bond-centered pairs rather than with respect to the creation of
single-particle excitations. This happens when the delocalization energy $%
-10J_{\mathrm{eff}}$ overcomes the energy $2(U_{2}-\mu )-2J_{\mathrm{eff}}$
needed to create a pair excitation. It is interesting to note that an
equivalent scenario does not happen for hole excitations, since hole
excitations decrease the occupation number and with that the tunneling
amplitudes.

To evaluate the boundary of the $\overline{n}=1$ Mott-insulator phase within
mean-field approximation, we construct a model for the excited bond-centered
pair excitations. When the number of pairs is small compared to the number
of sites, the Hamiltonian for the pairs living on top of a Mott state with
one particle per site can be written as
\begin{equation}
H_{\mathrm{pair}}=-J_{\mathrm{eff}}\sum_{\langle LL^{\prime }\rangle }\hat{p}%
_{L}^{\dagger }\hat{p}_{L^{\prime }}+2(U_{2}-\mu -J_{\mathrm{eff}})\sum_{L}%
\hat{n}_{L}^{p}.
\end{equation}%
Here, $L=\langle i,j\rangle $ labels the bonds of the cubic lattice and $%
\langle LL^{\prime }\rangle $ denotes pairs of nearest neighbors of these
bonds as they are described by the three-dimensional checkerboard lattice
(cf.\ Fig.~\ref{checker}).
Moreover, we have defined the bosonic creation and destruction operators for
bond-centered pair-excitations $\hat{p}_{L}^{\dagger }$ and $\hat{p}_{L}$,
with number operator $\hat{n}_{L}=\hat{p}_{L}^{\dagger }\hat{p}_{L}$. As a
consequence of the diluteness assumption, we have neglected the interaction
between pairs, arising if pairs occupy neighboring bonds. Since the
transition to a pair-superfluid will happen with the creation of a single
pair, this approximation will not influence the phase boundary. The energy
of a condensate of bond-centered pairs can now be estimated in a similar
fashion as before by making a product ansatz $\prod_{L}|\Phi _{p}\rangle _{L}
$ of coherent states being a superposition of zero and one pair at each
bond,
\begin{equation}
|\Phi _{p}\rangle _{L}=\cos \theta _{e}|0\rangle _{L}+\sin \theta
_{e}|1\rangle _{L}.  \label{ss1}
\end{equation}%
The order parameter of the pair condensate is defined by
$\langle \hat{p}_{L}\rangle =\frac{1}{2}\sin (2\theta _{e})$.
According to this ansatz, the
variational mean-field energy per site can be written as
\begin{equation}
\frac{E_{\mathrm{ep}}}{3N}=-\frac{z^{\prime }J_{\mathrm{eff}}}{4}\sin
^{2}2\theta _{e}+2(U_{2}-J_{\mathrm{eff}}-\mu )\sin ^{2}\theta _{e}
\label{ven1}
\end{equation}%
where $z^{\prime }=10$ is the coordination number of the three-dimensional
checkerboard lattice. The mean-field approach gives the same phase boundary
for the appearance of a pair condensate with finite order parameter $\langle
\hat{p}_{L}\rangle $ as the perturbation theoretical considerations of the
previous paragraph. The equivalence of both
approaches is generally given for an ansatz like (\ref{ss1}) which
includes only two states per site.

In Figure \ref{epp} we plot the results of minimizing $E_{\mathrm{ss}}, E_{
\mathrm{ep}}$ with respect to $\theta, \theta_e$ for $a_s/a=0.15$. The
stable Mott region with respect to single particle-hole excitation is given
by the interior of the black solid and dotted line characterized by
$\theta=0 $. On the right hand side of the blue dashed line in Fig.~\ref{epp}
one finds a region where $\min[E_{\mathrm{ep}}]<\min[E_{\mathrm{ss}}]$ with
$\theta_e \neq 0$. Thus, here the system is characterized by
$\langle p_L\rangle\neq 0$ and $\langle b_i\rangle=0$, i.e., the state is a
superfluid of extended pairs (ePSF).

Condensates of extended pairs have also been proposed in the context of
dimer models of reduced dimensions, describing frustrated magnets like SrCu$%
_{2}$(BO$_{3}$)$_{2}$ \cite{mila1}. By approximating triplet excitations as
hard-core bosons, the authors of Ref.~\cite{mila2} argue that for correlated
hopping these bosons can condense in pairs. Such pairing processes also bear
resemblance to molecular condensation due to Feshbach resonances in an
optical lattice \cite{stoof}.

We would like to point out that triple, quadruple or higher order excitations
do not play a dominant role. The effective tunneling matrix
element of such excitations will be very small since it appears in third or higher order 
perturbation theory only. Therefore inside a $\overline{n}=1$ phase, triple and higher
excitations cannot lower their energy efficiently by delocalization. We can, thus,
exclude a superfluid of triples or higher order objects. However, there is another 
possible and competitive scenario we would like to mention.
Instead of exciting a triple or quadruple, one can create a huge cluster of
extra particles, i.e., a big spatial domain with doubly occupied sites. In this case,
within each cluster, the energy of the additional particles (on top of the
$\overline{n}=1$ Mott background) is not lowered by delocalization, but rather by
the attractive interaction between them as it appears in second order perturbation
theory. In the bulk of such a cluster, this gives a binding energy of
$-6J_\text{eff}$ per extra particle. In comparison, in the pair superfluid each
particle can lower its energy by $J_\text{eff}$ because of binding and further by another
$5J_\text{eff}$ because of delocalization (i.e., Bose condensation). Accordingly,
in leading order a superfluid of bond-centered pairs on top of the $\overline{n}=1$
Mott insulator is equally favorable as a phase separated state with spatial domains
hosting either a Mott insulator of filling $\overline{n}=1$ or $\overline{n}=2$.
As a consequence, we cannot reliably exclude phase separation by means of simple
variational arguments.

Before moving on, let us briefly discuss another issue:
In this article, we are working in a situation with the chemical potential fixed rather than
the particle number. This approach is actually quite suitable for the description of
experiments with ultracold atoms, provided the atoms are trapped by a sufficiently
shallow potential. In such a situation, the local density approximation applies and
different regions in the trap correspond to different values of the chemical
potential. However, if the trap is too steep for the local density approximation to
be valid, it might introduce also new physics. Consider the following example: The
phase separated state described in the preceding paragraph might not be favored in
the homogeneous system. But, because it is energetically very close to the pair
superfluid, it can be favored already when a slight potential difference is
introduced, helping to form $\overline{n}=2$ Mott domains in the region of slightly
lower potential energy. Such a scenario can spoil the local density approximation
already for a very weak trapping potential.

\section{Superfluidity of local (site-centered) pairs}

\label{lps} Now, let us consider a regime that can be achieved if the
interaction strength $a_s/a$ is increased further. Considering again the $%
\overline{n}=1$ Mott insulator, for increasing interaction a site-centered
pair excitation, described by $|P_i\rangle$, eventually becomes more
favorable than the bond-centered excitations described by $|P_{\langle
ij\rangle}\rangle$. This happens when the ratio $U_3/U_2$ is reduced so much
that the potential energy $3U_3$ needed to create a pair of particles on the
same site equals the potential energy $2U_2$ required to create a pair of
particles on neighboring sites. Such a situation is possible as can be
derived from Eq.~(\ref{sig}). In the limit of large $V_0 \gg E_R$ and $a_s/a$
we can write $d(n)/d_0 \approx (gn_i)^{1/5}$ resulting in $3U_3-2U_2 \approx
-0.02U_0$. If $|3U_3-2U_2|$ becomes comparable to or smaller than $J_{2,2}$, a
bond centered pair excitation $|P_{\langle ij\rangle}\rangle$ can transform
to a site-centered pair excitation $|P_i\rangle$ by a single-particle
tunneling process described by the matrix element
$J_{\mathrm{pair }}=\sqrt{6}J_{22}$. In this regime, the pairs occupy the
lattice given by both the
sites and the bonds of the cubic lattice (see Fig.~\ref{checker}, center).
By delocalizing on this lattice, a pair can reduce its kinetic energy by
$12J_{\mathrm{pair}}$. As long as this energy is bigger than the kinetic
energy reduction $24J_{1,2}$ which two non-paired particles can achieve by
delocalization, the pair is stable towards breaking; this is the case for $%
J_{2,2}>\sqrt{3/2}J_{1,2}$. Thus, the binding mechanism of the pair is based
solely on the delocalization of its center of mass. At $|3U_3-2U_2|\approx 0$
(given, e.g., for $a_s/a\approx 0.21$ when $V_0/E_R\approx16$), the $%
\overline{n}=1$ Mott insulator becomes unstable with respect to pair
creation when $12J_{\mathrm{pair}}$ exceeds $3U_3-2\mu$.
It is fascinating to observe the emergence of exotic lattice geometries as
illustrated in Fig.~\ref{checker} as a consequence of pair creation.

\begin{figure}[ht]
\begin{center}
\epsfig{file=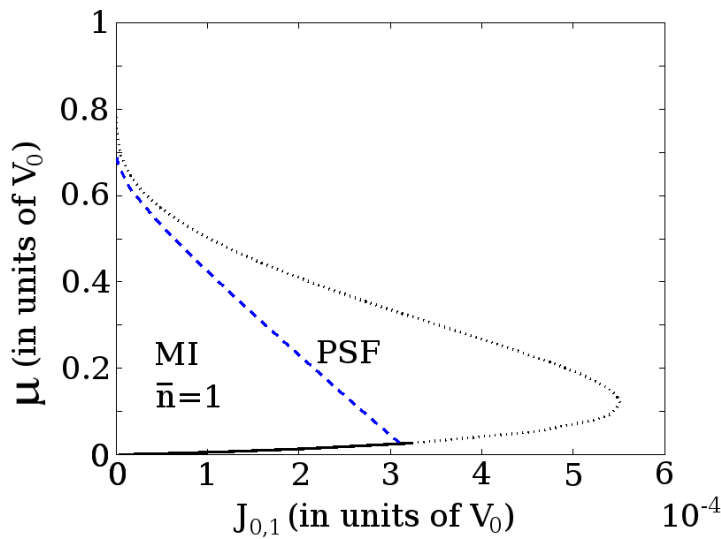,width=10cm}
\end{center}
\caption{Mott-insulator-to-superfluid phase transition for $a_s/a=0.3$. Inside the region
enclosed by the black solid and the blue dashed line, the system is a
Mott-insulator with $\overline{n}=1$ particles per site. Crossing the dashed blue
line, one enters a superfluid of local, site-centered pairs (PSF). Leaving the
Mott-phase by crossing the black solid line a superfluid of single particles
(or, equivalently, holes) is found. Black dashed line defined as in Fig.~\ref{epp}.}
\label{lpp}
\end{figure}

If the scattering length is increased further, such that $2U_2-3U_3\gg
J_{2,2}$, site-centered pair excitations $|P_i\rangle$ will be created
rather than bond centered ones $|P_{\langle ij\rangle}\rangle$. The
site-centered pair excitations can then tunnel from site to site coherently
via the occupation of a virtual bond-centered pair excitation. The
corresponding tunneling matrix element reads $J^{\prime}_{\mathrm{eff}}=6%
\frac{J_{2,2}^2}{2U_2-3U_3}=-J_{\mathrm{eff}}$. Moreover, the pair has a
binding energy of $6J^{\prime}_{\mathrm{eff}}$ (stemming from a small
perturbative admixture of the 6 neighboring bond-centered pair states).
Therefore, a site-centered pair is more favorable than two single-particle
excitations if $3U_3-12J^{\prime}_{\mathrm{eff}}<2(U_2-12J_{1,2})$.
If this condition is fulfilled, the Mott insulator becomes rather unstable
towards the creation of site-centered pair excitations than to the creation
of single particles. The instability occurs when $12J^{\prime}_{\mathrm{eff}%
} $ reaches $3U_3-2\mu$. 
As before, a mean-field calculation leads to the same phase boundary.
We plot the boundary of the $\overline{n}=1$ Mott phase for $a_s/a=0.3$ in
Fig.~\ref{lpp}. The instability towards the creation of single particles is
hardly important. It is predominantly the creation of single holes or
site-centered pairs of particles which destroys the Mott phase.

Note that in the limit of $U_{1}\gg J_{0,1}$ metastable repulsively bound
pairs of ultracold bosons have been observed in optical lattices \cite{wink,
petro}. Also, two-species mixtures of bosons with inter-species attraction
trapped in an optical lattice have been shown to give rise to superfluidity
of pairs \cite{kuk}. In the context of dipolar atoms in a two-leg ladder,
when no tunneling is present between the two legs, pair superfluidity arises
due to attraction between the dipolar atoms between the two legs of the
ladder \cite{san, lew}. Also, using a state-dependent optical lattice
potential, it is possible to create correlated tunneling of on-site pairs,
which in turn gives rise to superfluidity of local pairs \cite{rap, cir}. In
our present study, we find that such local pairing can emerge due to the
strong occupation-dependence of tunneling and on-site interaction.

After having studied the boundaries of the Mott-insulator phase with one
particle per site, let us have a look at the $\overline{n}=2$ Mott state. In
the limit of vanishing tunneling, a Mott state with two particles localized
at each site is favorable for $U_2<\mu <3U_3-U_2$. The upper border of this
interval is given by the potential energy difference of having three and two
particles at a site. This difference can, in fact, become lower than the
potential energy difference $U_2$ between two and one particle per site
marking the lower border. This is the case if $3U_3-2U_2<0$; then the $%
\overline{n}=2$ Mott-insulator phase is never stable with respect to the
creation of particle-hole pairs, irrespective of the tunneling strength; it
ceases to exist. The disappearance of the $\overline{n}=2$ Mott insulator
coincides with site-centered pair excitations becoming more favorable than
bond-centered ones in the limit of vanishing tunneling. Note that the
Mott-insulator phases with higher filling, $\overline{n}\ge3$, do not
disappear for large interaction $a_s/a$ within the Gaussian approximation.
The reason why these phases do not share the fate of the $\overline{n}=2$
Mott insulator is that the broadening on the Wannier-like site-wave
functions $\phi_i$ in response to adding one particle to that site becomes
less pronounced with increasing occupation: $U_2/U_3\ge U_3/U_4\ge U_4/U_5
\ge \cdots$. However, one should have in mind that for strong interaction,
sites occupied by three and more particles suffer strong dissipation due to
three-body collisions \cite{rem, zol}.

One might ask about the nature of the system's ground state at fixed filling
$n=2$ and for $3U_{3}-2U_{2}<0$, when there is no $\overline{n}=2$ Mott
phase. At vanishing tunneling, the ground state is highly degenerate
consisting of all Fock-states having occupation $n_{i}=1$ on half of the
sites and occupation $n_{i}=3$ on the others. Alternatively, one might say
that on top of a $\overline{n}=1$ Mott insulator, half of the sites are
occupied by additional site-centered pairs. For small but finite hopping
this degeneracy will be lifted. One can think of three possible scenarios: (i)
The pairs gather in one region in space; this corresponds to a phase
segregation between the $\overline{n}=1$ and the $\overline{n}=3$ Mott
phases. (ii) The pairs delocalize to form a superfluid. (iii) The pairs form
a checkerboard-type insulator avoiding pairs on neighboring sites. In order
to decide this question, we write down an effective Hamiltonian for the
site-centered pairs:
\begin{equation}
H_{\mathrm{pair}}=-J_{\mathrm{eff}}^{\prime }\sum_{\langle i,j\rangle
}c_{i}^{\dagger }c_{j}-\sum_{i}(2\mu -6J_{\mathrm{eff}}^{\prime
})n_{i}^{c}+(J_{\mathrm{eff}}^{\prime }-\Delta )\sum_{\langle ij\rangle
}n_{i}^{c}n_{j}^{c}
\end{equation}%
with bosonic pair annihilation and creation operators $\hat{c}_{i}$, $\hat{c}
_{i}^{\dag }$, and where we assume a hard-core constraint
$(\hat{c}_{i}^{\dag })^{2}=0$. The nearest-neighbor repulsion present in the last
term, with $\Delta =2\frac{J_{3,3}^{2}}{6U_{4}+U_{2}-6U_{3}}$, stems from
super-exchange processes between neighboring pairs. This model can be mapped
to a spin-1/2 XXZ model with the first term corresponding to the XX coupling
and the last one to the Z-coupling. Since $(J_{\mathrm{eff}}^{\prime
}-\Delta )\leq J_{\mathrm{eff}}^{\prime }$ is always true, the system will
neither form the checkerboard pattern (iii) (corresponding to an
antiferromagnetic state for the XXZ-magnet) nor show phase segregation (i)
\cite{bat}. The system forms a superfluid of site centered pairs (ii).

\section{Weakly interacting limit}

\label{bog} 
\begin{figure}[ht]
\begin{center}
\epsfig{file=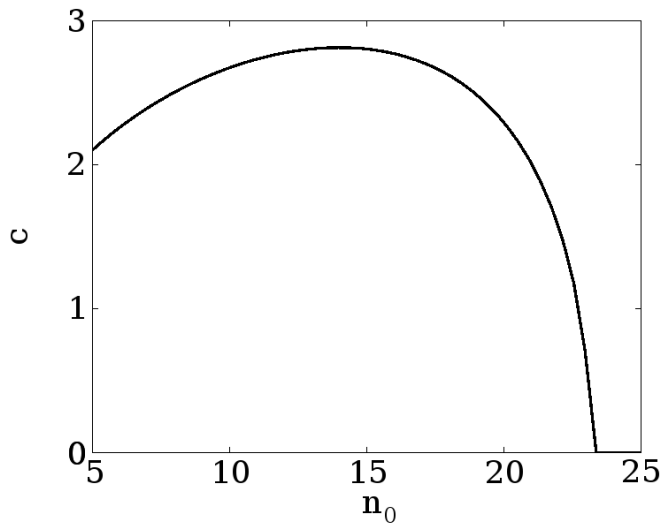,width=10cm}
\end{center}
\caption{Phonon velocity $c$ as a function of the superfluid occupation
number $n_0$. We find that after a critical occupation number, the phonon
velocity becomes imaginary, denoting a dynamical instability. The fixed
parameters are $a_s/a=0.01$ and $V_0/E_r=10$.}
\label{phon}
\end{figure}

Finally, we investigate the limit where interaction effects are
important not because of a large scattering length but because of large site occupation, i.e., $a_s/a \ll 1$, but the
mean number of particles per site $n_0\gg 1$. We assume small on-site number
fluctuations $\delta n \ll n_0$, i.e., $\sqrt{U_{n_0}/(n_0J_{n_0})} \ll 1$.
In this limit, we can write the modified Hubbard Hamiltonian as
\begin{eqnarray}  \label{mhub}
H &=& -J_{n_0}\sum_{ij}\hat{b}^{\dagger}_i[1+\alpha(\delta\hat{n}_i+\delta%
\hat{n}_j)]b_j
\nonumber\\&&
+\,\frac{U_{n_0}}{2}
\sum_i \hat{n}_i(\hat{n}_i-1)[1+\beta-2\beta(\hat{n}_i-1)]-\sum \mu
\hat{n}_i,
\end{eqnarray}
where
\begin{eqnarray}
\beta &=& \frac{3}{5}\sqrt{\frac{\pi}{2}} \left [ \frac{V_0}{E_R} \right
]^{1/4} \frac{a_s}{a}, \\
\alpha &=& \frac{\pi^{5/2}}{10\sqrt{2}} \left [ \frac{V_0}{E_R} \right
]^{3/4} \frac{a_s}{a}, \\
\frac{J_{n_0}}{V_0} &=& \left( \frac{\pi^2}{4}-1 \right) \exp \left [ -%
\frac{\pi^2}{4}\sqrt{\frac{V_0}{E_R}}
 \left [ 1-\frac{2\sqrt{2\pi}}{5}\left [ \frac{V_0}{E_R}
\right ]^{1/4} \frac{a_s}{a} n_0 \right ] \right ],
\end{eqnarray}
and $\delta\hat{n}_{i,j}=\hat{n}_i-n_0$. Here, we would like to point out
the similarity of Hamiltonian (\ref{mhub}) to the Quantum Ablowitz-Ladik (AL)
model for $q$-deformed bosons \cite{al}, given by
\begin{equation}  \label{alm}
H_{\mathrm{AL}}=-\sum_{i} [ B^{\dagger}_iB_{i+1} + B^{\dagger}_{i+1}B_i +
\frac{1}{2\gamma}\ln(1-QB^{\dagger}_iB_{i})],
\end{equation}
where $[B_i, B^{\dagger}_i]=\exp[-2\gamma N_i]$, and $Q=1-\exp[-2\gamma]$.
In the limit of $\gamma \rightarrow 0$ and $\gamma N_i \ll 1$, Eq.~(\ref{alm})
reduces to the occupation-dependent modified Hubbard model Eq.~(\ref{mhub})
with $\alpha=\gamma$ and $U_{n_0}=0$. It is found that in one and higher
dimension the AL model contains localized solutions \cite{cai, kev}. To
investigate this possibility, we first solve Eq.~(\ref{mhub}) in the
superfluid limit, where the order parameter reads
$\langle b_i \rangle=\sqrt{n_0}$. To look for fluctuations around the ground
state, we first convert the Hamiltonian in Eq.~(\ref{mhub}) to momentum space
by defining $b_i=\sum_i b_k \exp(-i\vec{k}.\vec{r}_i)$,
$\epsilon_k=4\sum_{i=1,2,3} \sin^2 \frac{k_ia}{2}$, and
$\gamma_k=4\sum_ {i=1,2,3} \cos^2\frac{k_ia}{2}$. Neglecting correlations
arising from the three-body interaction term in Eq.~(\ref{mhub}), one arrives at the
Hamiltonian
\begin{eqnarray}  \label{bham}
H_{\mathrm{mod}}&=&-\frac{n^2_0U_{n_0}}{2}+\sum_{k}J_{n_0} \epsilon_k
b^{\dagger}_k b_k
\\\nonumber&&
+\,\sum_{k}\left[ \frac{n_0U_{n_0}}{2}(1+\beta-2\beta (n_0-1))-\alpha
J_{n_0} n_0 \gamma_k \right]
\\&&\times
\left(2b^{\dagger}_k b_k+ b^{\dagger}_kb^{\dagger}_{-k} + b_k b_{-k}\right).
\end{eqnarray}
It can be diagonalized via a Bogoliubov transformation, and the excitation spectrum
$\Omega_k$ of the superfluid is found to be given by
\begin{equation}
\Omega^2_k=J^2_{n_0}\epsilon^2_k+2U_{n_0} n_0\left (1+\beta-2\beta
(n_0-1)-2\alpha \frac{J_{n_0}}{U_{n_0}} \gamma_k \right )\epsilon_k.
\end{equation}
In a cubic lattice, as $k\rightarrow 0$ one finds $\Omega_k/J_{n_0}U_0=c |k|a$,
where $c$ is the phonon velocity given by
\begin{equation}  \label{dyn}
c=\sqrt{(1+\beta-2\beta (n_0-1))-2\alpha \frac{J_{n_0}}{U_{n_0}} \gamma_0}.
\end{equation}
In Fig.~\ref{phon}, we plot the phonon velocity $c$ as a function of the
filling fraction $n_0$ for $a_s/a=0.01$. We find that initially, for
increasing $n_0$, the phonon velocity increases. But for higher $n_0$ the
phonon velocity starts decreasing due to the attractive effect of the
occupation dependent tunneling term, until the phonon velocity becomes
imaginary for a critical $n_0$. This results in a dynamical instability of
the superfluid when we are within the limit $\frac{a_s}{a}n_0 < 1$.
This instability occurs due to the attractive effect of the occupation
dependent tunneling, which can overcome the decreased repulsive on-site
interaction depending on the number of particles per site $n_0$. To
understand the effect of this instability, we first make a transition from
the discrete Hubbard model to a continuous model applicable for $ka \ll 1$
with a continuous field $\phi(r)$,
\begin{equation}
H_\mathrm{cont}=-\int d^3r \phi^*(r)\nabla^2 \phi(r) + \frac{U}{2}\int V_{\mathrm{eff}}
	(r-r^{\prime})|\phi(r)|^2|\phi(r^{\prime})|^2.
\end{equation}
Here, the distance is expressed with respect to the lattice constant $a$,
and the effective interaction potential is given by $V_{\mathrm{eff}
}(r-r^{\prime})=\mathcal{F}^{-1}[1+\beta-\beta(n_0-1)-2\alpha\frac{J_{n_0}}{
U_{n_0}}\gamma_k]$, where $\mathcal{F}^{-1}$ stands for the inverse Fourier
transformation. Using a Gaussian ansatz along one direction, say $x$, and
uniform in the other directions, $\phi(r)=1/\pi^{1/4}d^{1/2}_s
\exp(-x^2/2d^2_s)$, the energy functional for the self-trapped state reads
$E_{\mathrm{sol}}=1/d^2_s+\frac{U_{n_0}}{J_{n_0}\sqrt{2\pi}}
\left(1+\beta-2\beta(n_0-1)-\alpha\frac{2J_{n_0}}{U_0}(5+\exp(-2/d^2_s))
\right)/d_s.$ When $n_0$ exceeds a critical density,
$E_{\mathrm{sol}}$ is minimized for a finite $d_s \gg 1$. Thus, the homogeneous
superfluid becomes dynamically unstable towards a state which is localized
only in one direction, forming a 2D slab.

\section{Conclusion and Outlook}

In this paper we have predicted various effects resulting from
interaction-induced band mixing in systems of ultracold bosonic atoms in optical
lattice potentials. We have derived the modified bosonic Hubbard model (\ref{hub2})
having occupation-number-dependent parameters. This model comprises an effective
interaction-induced broadening of the Wannier-like single-particle orbitals, and,
thus, captures also the situation when the s-wave scattering length becomes
comparable to the lattice spacing, $a_{s}/a\rightarrow 1$. Using this model, we
find that for scattering lengths $a_{s}\sim 0.15a$ and lattice depths
$V_{0}\sim 12E_{R}$, the $\overline{n}=1$ Mott-insulator state can become unstable
towards a superfluid which consists of bond-centered pair excitations.
This scenario is novel considering the fact that the extended pairs emerge due to
the occupation dependence of both the tunneling strength and the
on-site interaction. For even higher interaction, the nature of the superfluid pair
excitations (destroying the insulator) changes. The pairs can now occupy both the
bonds on the lattice (i.e., two neighboring sites) or its sites; in that way an
exotic lattice geometry as shown in the center plot of Fig.~\ref{checker} emerges.
Increasing the interaction further, eventually the pairs live only on
the sites of the lattice. In this regime of high interaction strength, the $\overline{n}=2$ Mott state gets
completely destroyed by the site-centered pair fluctuations.
We have also looked into the regime where interaction induced Wannier-broadening
arises from large filling $\overline{n}\gg 1$ at small scattering lengths,
$a_{s}\ll a$. In this limit, we found that the superfluid becomes dynamically
unstable due to the attractive nature of the occupation-dependent tunneling.
The system then transforms from a uniform superfluid state to an asymmetric state
which is localized in one direction and extended in the other two directions.

In future studies, we would like to study the role of dissipation in these systems.
Also, a more accurate determination of the number dependence of the Hubbard
parameters $J_{n_{i},n_{j}}$ and $U_{n_{i}}$ will be required for a quantitative
description of the effects described here. Finally, it would also be worth
studying in detail the role of a trapping potential, as it is present in experiments.

\section{Acknowledgments}

This work is financially supported by Spanish MICINN (FIS2008-00784 and
Consolider QOIT), EU Integrated Project AQUTE, ERC Advanced Grant QUAGATUA,
the EU STREP NAMEQUAM, the Caixa Manresa, the Alexander-von-Humboldt
foundation, and the German-Israel Foundation (grants 149/2006 and
I-1024-2.7/2009). B.A.M. appreciates hospitality of the Institut de Ci\`{e}%
ncies Fot\`{o}niques (Barcelona, Spain)

\end{document}